\documentclass[a4paper,11pt]{article}
\usepackage[dvips]{graphicx}
\usepackage{amssymb}
\usepackage{amsmath}

\newcommand{\plb}{Phys. Lett. B. }

\newcommand{\arx}{arXiv:}

\begin{document}

\title{Slab Bag Fermionic Casimir effect, Chiral Boundaries and Vector Boson - Majorana Fermion Pistons}
\author{
V.K.Oikonomou $^{1,\star}$
\thanks{voiko@physics.auth.gr}\\
\emph{$^{\star}$Dept. of Theoretical Physics Aristotle University of Thessaloniki},\\
\emph{Thessaloniki 541 24, Greece}\\
\emph{$^{1}$Technological Education Institute of Serres, 62124 Serres, Greece} \\
and \\
N.D.Tracas
\thanks{ntrac@central.ntua.gr}\\
\emph{Physics Department, National Technical University of Athens},\\
\emph{Athens 157 73, Greece}
}
\maketitle

\begin{abstract}
In this article we consider the Casimir energy and force of
massless Majorana fermions and vector bosons between parallel
plates. The vector bosons satisfy perfect electric conductor
boundary conditions while the Majorana fermions satisfy bag and
chiral boundary conditions. We consider various piston
configurations containing one vector boson and one fermion. We
present a new regularization mechanism the piston offers. In our
case regularization occurs explicitly at the Casimir energy
density and not at the Casimir force level, as in usual pistons.
We make use of boundary broken supersymmetry to explain the number
of fields that appear in all the studied cases. The effect of
chiral boundary conditions in a fermion boson system is
investigated. Concerning the supersymmetry issue and the vanishing
Casimir energy, we study a massive Dirac fermion-scalar boson
system with bag and specific Robin type boundary conditions for
which the Casimir energy vanishes. Finally a two scalar bosons
system between parallel plates is presented in which the
singularities vanish in the total Casimir energy. A discussion on
boundary broken supersymmetry follows.
\end{abstract}

\bigskip

\section*{Introduction}

One of the most interesting and most studied phenomena in Quantum
Field Theory is the Casimir effect. It is a quantitative proof of
a quantum field quantum fluctuation.
It originates from the ``confinement" of a field in finite volume and
many studies have been devoted on this  phenomenon  since H. Casimir's original
work \cite{Casimir}. The Casimir energy, is closely related to the
boundary conditions of the fields under consideration which modify
the nature of the so-called Casimir force
generated
by the vacuum energy. Casimir calculated the
electromagnetic force between two parallel  conducting  plates which he found
to be attractive. A repulsive Casimir force,
in the case of a conducting sphere,
was calculated by Boyer \cite{boyer} some time later,

\noindent Of course the attractive and repulsive nature of the
Casimir force has many applications in nanotubes and
nanotechnology since a collapsing force can lead to the
destruction of such a configuration.
Therefore, the stabilization of such a system is highly important.
The study was generalized
to include other (apart from electromagnetic) quantum fields such as fermions, bosons and other
scalar fields (see for example \cite{Bordagreview,elizalde} and
references therein). The boundary conditions modify the Casimir
force for all the quantum field cases. The most used ones are
Dirichlet and Neumann boundary conditions on the plates however
this is not the case for fermion quantum fields. Dirichlet and
Neumann boundary conditions have no direct generalization in the
case of fermion fields and in general for fields with spin$\neq 0$
\cite{ambjorn}. In that case the bag boundary conditions are used.
These boundary conditions, in the case of fermion fields, were
introduced to provide a solution to confinement \cite{bag}.

\noindent In this paper we shall extensively use the bag boundary
conditions (and their modified form known as the chiral bag boundary
conditions) for a Majorana fermion field confined between two
parallel plates and in piston configurations with the
aforementioned boundary conditions.

\noindent The extension of the two slabs to pistons is motivated
mainly from the regularization that the Casimir piston offers. As is well
known, the Casimir energy contains singularities that must be
regularized in order to acquire a finite result
There are two ways to compute the Casimir energy, the
cutoff method and the zeta function regularization method. In the
former case the singularities are regularized by
introducing suitable counter-terms that cancel the singularities.
The Casimir piston configuration offers a very elegant way of
cancelling these singularities. This configuration was originally
used \cite{calvacanti} as a single rectangular box with three
parallel plates where the middle one is free to move. The
dimensions of the piston are $(L-a)\times b$ and $a\times b$, with
the moving plate being located in $a$. In \cite{calvacanti} the
Casimir energy and Casimir force for a scalar field was calculated
for a piston. The boundary conditions on the ``plates'' where the
Dirichlet ones. The literature on the subject is quite big,
\cite{KirstenKaluzaKleinPiston,ElizaldePistons,KirstenPistons,fulling,Cheng,edery1,ar1,ar2,ar3,Teo,teo2,teo1,oikonomou}
calculating the Casimir force for various configurations of the boundary
conditions of the scalar field, in both massive and massless case.
The regularization the Casimir piston provides is very useful.
Actually, when one calculates the Casimir energy between parallel
plates confronts, as we mentioned, infinities that must be
regularized. The regularization of the Casimir energy in the
parallel plate geometry can be performed if we calculate it as a
sum over discrete modes (due to boundary conditions on the plates)
minus the continuum contribution (plates distance send to
infinity)\cite{Bordagreview,elizalde}. The discrete sum consists
of three terms, a volume divergent one (which is cancelled by the
continuum integral), a surface divergent one and a finite term.
This can be easily seen if the calculations of the Casimir energy
are performed with the introduction of a UV cutoff. Before the use of
the piston, the surface divergent term was
thrown out ``by hand''; a completely unjustified action since that term
cannot be removed by renormalizing the physical parameters of the theory \cite{jaffenew}.
On the other hand,
the zeta function regularization technique renormalizes this term
to zero. Thus the cutoff technique and the zeta regularization
technique agree perfectly.
The Casimir piston solves this problem in a very elegant way,
because the surface terms of the two piston chambers cancel each other and
thus the Casimir force can be calculated in a consistent way
\cite{KirstenKaluzaKleinPiston,ElizaldePistons,KirstenPistons,fulling,Cheng,edery1,ar1,ar2,ar3,Teo,teo2,teo1,oikonomou}.

\begin{figure}[t]
\begin{center}
\includegraphics[scale=.7]{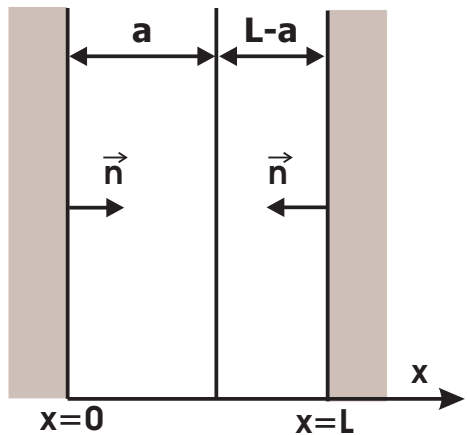}
\end{center}
\caption{The Casimir piston in $x$ dimension} \label{olaskata}
\end{figure}

\noindent In this article we shall impose bag boundary conditions
for Majorana fermions between two slabs and construct, in some
cases, a piston of such slabs (see (\ref{olaskata})).
 Since the bag boundary conditions confine the fermion
field in one of the two piston chambers, it is supposed that every
piston chamber contains a different fermion flavor so no theoretical
inconsistency occurs. We should mention that
the piston with bag boundary condition fermions
is in one dimension (while the other dimensions are infinite),
since the confinement of a half integer spin field can be implemented only in this case
(the presence of corners prevents solutions to the massless Dirac
equation; see \cite{paola} and references therein). The massive
fermion field is discussed in \cite{bag1,elizaldeslab}.

\noindent The slab fermion Casimir effect for a piston shall be
also studied. It is always assumed that fermions appearing in
different chambers must have different flavors, for consistent
slab boundary conditions. The fermion piston has nothing new to
offer as far as regularization method is concerned but we shall compute the Casimir
force in order to check the validity of the rules that hold in the boson piston case.
It would be interesting to see that the fermionic Casimir force between
plates is attractive, contrary to what one expects.

\noindent
After the simple Majorana fermion piston, we shall
examine the Majorana-fermion and vector-boson chamber and piston. The
connection with supersymmetry shall be examined. In the case of
Majorana-fermion and vector-boson (in the following fermion refers to
Majorana fermion and boson to vector boson unless stated
differently) piston we shall study three different cases, namely,
fermion in one chamber ($a$-chamber) and boson in the
other chamber ($L-a$-chamber), boson in the $a$-chamber and
fermion in the $L-a$-chamber, and finally boson-fermion in the
$a$-chamber and boson-fermion in the $L-a$-chamber (with different fermion
flavors).

\noindent  The motivation to use a piston with one chamber filled
with a fermion and the other with boson, is even stronger. This
comes from the fact that the Casimir energy density is
regularized, and when the $L$ parameter is send to infinity, the
Casimir energy yields the vector boson or Majorana fermion Casimir
energy, of course without singularity. Thus this could be another
regularization use of the piston with the new feature of using
Majorana fermions and vector bosons. This could be particularly
interesting since we can use a fermion just to regularize the
vector boson field (photon) Casimir energy and then forget it. The
new result within these considerations is that Casimir energy
density (and therefore the Casimir energy) is regularized and not
just the force.
The motivation to study fermion boson chamber is the
automatic cancelation of the singularities when the number of
fermions and bosons follow the vector superfield rules dictated
by supersymmetry. Of course supersymmetry is broken due to
boundary conditions but a remnant of supersymmetry can be really
useful.

\noindent We shall extend the boson fermion chambers and pistons
in the case where the fermion obeys chiral boundary conditions on
the plates. In that case the minimization of the energy as a
function of the chiral parameter $\phi$ is examined. Also the sign
of the Casimir force as a function of $\phi$ is also studied.
Our next step will be to introduce two real scalar bosons, obeying
Robin boundary conditions, and a massive Dirac fermion obeying bag boundary conditions.
Although the number of the fields are determined by the supersymmetric chiral superfield rules,
supersymmetry is broken due to boundary conditions.
In that case the total Casimir energy can be zero in some cases

\noindent Finally we shall examine the massive scalar bosons
Casimir effect between parallel plates in the case that mixed
boundary conditions are imposed.
The connection of this case with the rest of the present study is that the
singularities cancel also. Indeed if one uses a
boson, satisfying Dirichlet-Neumann boundary conditions on the
plates, together with a scalar boson, satisfying Dirichlet-Dirichlet
boundary conditions, the singularities of the Casimir energies for each case
cancel each other.

\noindent Finally, we present the conclusions and we discuss
some applications of the above study.

\section{Majorana Fermions in a Slab and Bag Boundary\\ Conditions}

Consider Majorana fermion fields in 3 space dimensions
while one dimension is considered finite by two parallel plates
at $x=0$ and $x=L$.
We shall also assume that fermions are not allowed to exist outside the
parallel plate system. This is actually the MIT bag boundary
conditions which are expressed as
\begin{equation}\label{mitbag}
in^{\mu}\gamma_{\mu}\psi=\psi
\end{equation}
or in Lorentz covariant form
\begin{equation}\label{mitbagcov}
n^{\mu}\bar{\psi}\gamma_{\mu}\psi=0
\end{equation}
which $n^{\mu}=(0,\vec{n})$ and $\vec{n}$ the vector normal to the
surface of the plates and directed to the interior of the slab
configuration. The above two equations show that there is no
fermion current flowing outwards of the parallel plates.

\noindent In most of the calculations we shall perform in this
article we shall use the cutoff method \cite{Bordagreview,bag1} in
order to compute the Casimir energy, since we want to see
explicitly the singularities that the Casimir energy has. The
eigenvalues of the massless Dirac equation obeying bag boundary
conditions are
\begin{equation}\label{dhyf}
\omega_n=\sqrt{k_T^2+\frac{\pi^2(n+\frac{1}{2})^2}{L^2}}
\end{equation}
where $k_T$ refer to the transverse components of the
momentum. Using the cutoff method the Casimir energy density per
unit area reads \cite{bag1},
\begin{equation}\label{casimiren}
\frac{\mathcal{E}_{f}}{L}=-\frac{1}{2L}2\sum_{n=0}^{\infty}\int\frac{\mathrm{d}^2k_T}{(2\pi)^2}{\,}\omega_n{\,}e^{-\tau
\omega_n}
\end{equation}
where $\mathcal{E}_f$ is the total Casimir energy per unit area between the plates
and we have introduced an exponential regulator in terms of the
cutoff $\tau$. The factor 2 for the Majorana
fermion comes from the spin and the degrees of freedom in four
spacetime dimensions. Using
\begin{equation}\label{com1}
\omega e^{-\omega \tau}=\big{(}\frac{\partial}{\partial
\tau}\big{)}^2\big{(}\frac{1}{\omega}e^{-\omega \tau}\big{)}
\end{equation}
and carrying out the integration we obtain
\begin{equation}\label{com2}
\frac{\mathcal{E}_{f}}{L}=-\frac{2}{L}\big{(}\frac{\partial}{\partial
\tau}\big{)}^2\frac{1}{2\pi \tau}\Big{(}\frac{e^{-\frac{\pi
\tau}{2L}}}{1-e^{-\frac{\pi \tau}{L}}}\Big{)}
\end{equation}
In the limit $\tau \rightarrow 0$ we obtain
\begin{equation}\label{before1}
\frac{\mathcal{E}_{f}}{L}=-\frac{7}{8}\frac{\pi^2}{720{\,}L^4}-\frac{3}{\pi^2\tau^4}
\end{equation}
where we can see explicitly the singularity in terms of the cutoff
$\tau$. Note that the series over $\tau$ contains only even
powers of $\tau$. This is a characteristic of fermions and vector
bosons. Consistent boundary conditions for massless vector bosons
are the so-called perfect conductor boundary conditions on the
plates, namely
\begin{equation}\label{comr1}
n_{\mu}F_{a}^{\mu {\,}\nu}=0
\end{equation}
Working in the same way as in the fermion case, the Casimir energy
density per unit area of a vector field satisfying perfect
conductor boundary conditions on the plates reads
\begin{equation}\label{bosoncut}
\frac{\mathcal{E}_{b}}{L}=\frac{1}{2L}\Big{(}2\sum_{n=1}^{\infty}\int\frac{\mathrm{d}^2k_T}{(2\pi)^2}{\,}\omega_{nb}{\,}e^{-\tau
\omega_{nb}}+\int\frac{\mathrm{d}^2k_T}{(2\pi)^2}{\,}|k_T|{\,}e^{-\tau
|k_T|}\Big{)}
\end{equation}
where $\omega_{nb}=\sqrt{k_T^2+\frac{\pi^2n^2}{L^2}}$. Upon
integration we obtain
\begin{equation}\label{comvb2}
\frac{\mathcal{E}_{b}}{L}=-\frac{1}{L}\big{(}\frac{\partial}{\partial
\tau}\big{)}^2\frac{1}{2\pi \tau}\Big{(}\frac{1+e^{-\frac{\pi
\tau}{L}}}{1-e^{-\frac{\pi \tau}{L}}}\Big{)}
\end{equation}
and in the limit $\tau \rightarrow 0$ we obtain
\begin{equation}\label{8before}
\frac{\mathcal{E}_{b}}{L}=\frac{3}{\pi^2\tau^4}-\frac{1}{720}\frac{\pi^2}{L^4}
\end{equation}
where $\mathcal{E}_{b}$ is the total Casimir energy per unit area
of the boson between plates. We shall make extensive use of the
above well known expressions in the rest of the paper (for a
similar analysis see \cite{muta}). Note again that the series over
$\tau$ contains even powers only, just as in the fermion case. For
completeness we shall present briefly the scalar boson case. The
scalar boson Casimir energy density reads
\begin{equation}\label{scalarbosoncut}
\frac{\mathcal{E}_{\mathrm{scalar}}}{L}=\frac{1}{2L}\sum_{n=1}^{\infty}\int\frac{\mathrm{d}^2k_T}{(2\pi)^2}{\,}\omega_{nb}{\,}e^{-\tau
\omega_{nb}}
\end{equation}
and after integrating, we obtain
\begin{equation}\label{bosonscalar}
\frac{\mathcal{E}_{\mathrm{scalar}}}{L}=\frac{1}{L}\big{(}\frac{\partial}{\partial
\tau}\big{)}^2\frac{1}{\pi \tau}\Big{(}\frac{1}{-1+e^{\frac{\pi
\tau}{L}}}\Big{)}
\end{equation}
In the limit $\tau \rightarrow 0$, the Casimir energy density for
the scalar field reads
\begin{equation}\label{8bef56}
\frac{\mathcal{E}_{\mathrm{scalar}}}{L}=\frac{6}{\pi^2\tau^4}-\frac{1}{L\pi\tau^3}-\frac{1}{720}\frac{\pi^2}{L^4}
\end{equation}
It is obvious, contrary to the fermion and vector boson case, the
scalar boson energy density contains even and odd powers of
$\tau$.

\section{Different Flavor Fermionic Casimir Piston}

Let us see how the piston configuration that we mentioned in the
introduction helps in the cancellation of the singularities.
Consider a piston with two chambers that are constructed by three
parallel plates at $x=0$, $x=a$ and $x=L$. The total Casimir
energy per unit area of a Majorana fermion is equal to,
\begin{equation}\label{toto1}
\mathcal{E}_{piston}=\mathcal{E}_{f}(a)+\mathcal{E}_{f}(L-a)
\end{equation}
and relation (\ref{before1}) gives
\begin{equation}\label{toto2}
\mathcal{E}_{piston}=-\frac{3a}{\pi^2\tau^4}-\frac{1}{720}\frac{\pi^2}{8{\,}a^3}-\frac{3(L-a)}{\pi^2\tau^4}-\frac{1}{720}\frac{\pi^2}{8{\,}(L-a)^3}
\end{equation}
Taking the derivative of the total Casimir energy per unit area we
obtain the Casimir force  per unit area
\begin{equation}\label{fp}
F_c=-\frac{\partial \mathcal{E}_{piston}}{\partial a}
\end{equation}
which is equal to
\begin{equation}\label{toto3}
F_c=-\frac{3}{\pi^2\tau^4}-\frac{3}{720}\frac{\pi^2}{8{\,}a^4}+\frac{3}{\pi^2\tau^4}+\frac{3}{720}\frac{\pi^2}{8{\,}(L-a)^4}
\end{equation}
In the above relation it is clear that the total Casimir force per
unit area is finite.

\noindent
Relation (\ref{toto3}) is the Casimir force on a piston plate in
the case the chambers are filled with different flavors of
fermions (in order bag conditions are consistently fulfilled).
Vacuum fluctuations of massless fermions between two
parallel and confining plates give rise to an attractive Casimir
force. Thus both vector bosons and fermions lead to a negative
Casimir force, contrary to what would be expected from fermions
due to Fermi-Dirac statistics. The Casimir force behaves
exactly as in the bosonic case. In detail the
force is attractive when $a$ is small and repulsive when $L-a$ is
small (see also
\cite{KirstenKaluzaKleinPiston,ElizaldePistons,KirstenPistons,fulling,Cheng,edery1,ar1,ar2,ar3,Teo,teo2,teo1}).
Finally, it is obvious that in the limit $L\rightarrow \infty$ we
obtain the usual Majorana fermionic Casimir force between plates
(see \cite{paola,elizaldeslab}).

\section{Majorana Fermion-Vector Boson Chambers and Piston Configurations}

Consider massless, non interacting vector bosons and Majorana
fermions, described by the Lagrangian,
\begin{equation}\label{lagr}
\mathcal{L}=-\frac{1}{4}F_{\mu{\,}\nu}F^{\mu{\,}\nu} +i\overline{\Psi }%
\gamma ^{\mu }\partial _{\mu }\Psi ,
\end{equation}%
We shall assume that the vector boson field and the Majorana
fermion co-exist between two parallel plates located at $x=0$ and
$x=L$. Thus we construct a chamber filled with bosons and fermions
in a ``supersymmetric" way. Indeed Lagrangian (\ref{lagr}) looks
like the $N=1$, $d=4$ vector supersymmetric Lagrangian describing
photons and the supersymmetric partner, the photino. We should
clarify at this point that
supersymmetry is broken when the bag boundary conditions
for fermions and perfect conductor boundary
conditions are used for bosons on the plates.
Thus in this physical system,
supersymmetry is explicitly  broken. A similar effect happens
when physical systems are studied at finite temperature, where
supersymmetry breaks due to
different boundary conditions applied to bosons and fermions.
However in the slab case
there exists a remnant of supersymmetry. In fact
we shall use the above Lagrangian in order to determine formally
the appropriate number of fermions and bosons fields we should use
for singularity cancellation.

\subsection{Massless Vector Boson-Majorana Fermion Chamber}

Consider one vector boson and one Majorana fermion field described
by the Lagrangian (\ref{lagr}). We assume that the fields are
confined between two parallel plates on which fermions satisfy bag
boundary conditions and bosons satisfy perfect conductor, that is
\begin{equation}
n^{\mu}\bar{\Psi}(x=0)\gamma_{\mu}\Psi(x=0)=0,\qquad\qquad{\,}n^{\mu}\bar{\Psi}(x=L)\gamma_{\mu}\Psi(x=L)=0
\end{equation}
for fermions and
\begin{equation}
n_{\mu}F_{a}^{\mu {\,}\nu}=0
\end{equation}
for vector bosons. The total Casimir energy per unit area of the
slab system reads (Casimir slab chamber)
\begin{equation}\label{15}
\mathcal{E}_{\mathrm{total}}=-\frac{1}{720}\frac{\pi^2}{L
^3}+\frac{3L}{\pi^2\tau^4}-\frac{7}{720}\frac{\pi^2}{8L^3}-\frac{3L}{\pi^2\tau^4}
\end{equation}
It is obvious that the Casimir energy per unit area for the plate
chamber is singularity free, since the singularity of the fermion
energy is canceled by the corresponding ones of the two bosons. This
is a very valuable result and it is due to the
supersymmetry remnant that the system possesses. The total Casimir force per
unit area between the plates is
\begin{equation}\label{16}
F_{\mathrm{total}}=-\frac{21}{720}\frac{\pi^2}{8L^4}-\frac{3}{720}\frac{\pi^2}{L^4}
\end{equation}
which, as we can see, is attractive. Thus, when we consider the
fermion boson chamber, the force between the plates is even more
attractive compared to the single boson (fermion) case. It is not
difficult to extend the chamber to a piston configuration, just to
see how the force behaves. Of course each chamber is singularity
free. Thus adding the contributions from the two chambers, $a$ and
$L-a$, we obtain, the total Casimir energy (it is assumed that each
chamber contains \emph{different} fermion flavor)
\begin{equation}\label{17}
\mathcal{E}_{\mathrm{total}}=-\frac{1}{720}\frac{\pi^2}{\alpha
^3}-\frac{7}{720}\frac{\pi^2}{8\alpha
^3}-\frac{1}{720}\frac{\pi^2}{(L-\alpha
)^3}-\frac{7}{720}\frac{\pi^2}{8(L-\alpha )^3}
\end{equation}
and the total Casimir force per unit area is given by
\begin{equation}\label{18}
F_{\mathrm{total}}=-\frac{21}{720}\frac{\pi^2}{8\alpha
^4}-\frac{3}{720}\frac{\pi^2}{\alpha^4}+\frac{21}{720}\frac{\pi^2}{8(L-\alpha
)^4}+\frac{3}{720}\frac{\pi^2}{(L-\alpha )^4}
\end{equation}
Notice that in the limit $L\rightarrow \infty$ relation (\ref{18})
becomes identical to relation (\ref{16}). Thus we recover the
initial system, as is expected in every piston configuration.

\section{Massless Boson-Fermion Piston.\\
A Regularization Method of the Casimir Energy Density per
Unit Area Using the Piston}

As we mentioned in the introduction, the most theoretically
attractive feature of the piston is that it renders the Casimir
force of a bosonic (and fermionic) system finite. However the
Casimir energy density (and therefore the total Casimir energy)
for the aforementioned fields, even for the piston configuration,
contain singularities. We shall demonstrate a use that the Casimir
piston offers to the regularization of the Casimir energy density. We
will make use of the Casimir piston configuration but we shall
fill the two chambers with different statistics fields. To be specific,
consider that the $a$ chamber is filled with a Majorana
fermion with bag boundary conditions on the boundaries while the
$L-a$ chamber is filled with a vector boson satisfying perfect conductor
boundary conditions on the two plates.
Notice that the system as a whole is determined by the Lagrangian (\ref{lagr}), so there is a
remnant supersymmetry, which is however broken since the fields
are confined in different places in space, and of course the
boundary conditions are different. The fermionic Casimir energy
energy density per unit area is equal to
\begin{equation}\label{7}
\mathcal{U}_f(\alpha)=-\frac{7}{720}\frac{\pi^2}{8\alpha^4}-\frac{3}{\pi^2\tau^4}
\end{equation}
while the vector boson has Casimir energy density
\begin{equation}\label{8}
\mathcal{U}_b(L-a)=\frac{3}{\pi^2\tau^4}-\frac{1}{720}\frac{\pi^2}{(L-\alpha
)^4}
\end{equation}
The total Casimir energy density for the piston system is
$\mathcal{U}_{total}=\mathcal{U}_f(a)+\mathcal{U}_{b}(L-a)$.
Adding the above two contributions notice that the total Casimir
energy density is finite, and equal to
\begin{equation}\label{8p}
\mathcal{U}_{total}=-\frac{1}{720}\frac{\pi^2}{(L-\alpha
)^4}-\frac{7}{720}\frac{\pi^2}{8\alpha^4}
\end{equation}
In the above relation, the total Casimir energy density behaves as
follows,
\begin{equation}\label{totalfunction}
\mathcal{U}_{total}=\bigg{\{}%
\begin{array}{c}
 \mathcal{\bar{U}}_f \quad\textrm{for}\quad 0\leq x \leq a\\
  \mathcal{\bar{U}}_b \quad\textrm{for}\quad a\leq x \leq L\\
\end{array}%
\end{equation}
where $\bar{\mathcal{U}}_f$ is the regularized Majorana
fermion energy density and $\bar{\mathcal{U}}_b$ is the
corresponding one for the vector boson. Therefore,
the energy density in each chamber is made regular and, of course, the total
energy density is regular. The
total fermionic Casimir energy density per unit area is equal to
\cite{fulling_b},
\begin{equation}\label{fulling1}
E_f(a)=\int_{0}^{a}\bar{\mathcal{U}}_f(a)\mathrm{d}x=-\frac{7}{720}\frac{\pi^2}{8\alpha^3}
\end{equation}
since the fermionic energy density is regularized and independent
of $x$. In the same way the vector boson total Casimir energy
density per unit area is,
\begin{equation}\label{fulling2}
E_b(L-a)=\int_{a}^{L}\bar{\mathcal{U}}_b(a)\mathrm{d}x=-\frac{1}{720}\frac{\pi^2}{(L-\alpha
)^3}
\end{equation}
Adding the two bosons and fermion contributions we obtain the
total Casimir energy of the piston
\begin{equation}\label{9}
\mathcal{E}_{\mathrm{total}}=-\frac{1}{720}\frac{\pi^2}{(L-\alpha
)^3}-\frac{7}{720}\frac{\pi^2}{8\alpha^3}
\end{equation}
We can clearly see that the total Casimir energy is free of
singularities since we regularized the Casimir energy density. By
taking the limit $L\rightarrow \infty$ the total Casimir energy
becomes equal to the single Majorana fermion Casimir energy. Thus
the piston offers a regularization method for the fermion Casimir
energy (and more importantly for the vector boson). This was the
motivation to study such configurations. Of course the total
Casimir force on the piston plate is regularized and free of
singularities. Indeed, since $F=-\frac{\partial E}{\partial a}$,
taking the derivative of (\ref{9}) with respect to $a$, we obtain,
\begin{equation}\label{10}
F_{\mathrm{total}}=\frac{3}{720}\frac{\pi^2}{(L-\alpha
)^4}-\frac{21}{720}\frac{\pi^2}{8\alpha^4}
\end{equation}
Again in limit $L\rightarrow \infty$ we render the usual fermionic
Casimir force between plates with bag boundary conditions, divided
by 2 due to the Majorana degrees of freedom
\cite{paola,elizaldeslab,ravdal,ravdal1}.

\noindent The most important case comes when in the above piston
setup the fields are confined in the chambers with inverse order,
that is, vector boson in the $a$ chamber and Majorana fermions in
the $L-a$ chamber. Following the same procedure as in the
previous, we obtain the total Casimir energy
\begin{equation}\label{13}
\mathcal{E}_{\mathrm{total}}=-\frac{1}{720}\frac{\pi^2}{\alpha
^3}-\frac{7}{720}\frac{\pi^2}{8(L-\alpha )^3}
\end{equation}
and the total Casimir force,
\begin{equation}\label{14}
F_{\mathrm{total}}=\frac{21}{720}\frac{\pi^2}{8(L-\alpha
)^4}-\frac{3}{720}\frac{\pi^2}{\alpha^4}
\end{equation}
It is obvious that the total Casimir energy is again singularity
free, as was in the previous case. The difference of the two cases
is that in the limit $L\rightarrow \infty$, the total Casimir
energy becomes the Casimir energy of a massless vector boson field
confined between two plates. Thus the inverse piston configuration
serves as a regularization technique for the vector boson Casimir
energy. Of course the force is finite, as we can see (\ref{14}). The
new regularization that the piston offers within the setup we
presented is that the total energy is finite and not just the
force.

\noindent In conclusion, the piston can offer a regularization
method for the vector boson and Majorana fermion Casimir energy
when we make use of a ``supersymmetric" system in which fermions
and bosons are put in different chambers. Although supersymmetry
is broken due to boundary conditions, the singularities cancel in
a supersymmetric way, and the Casimir energy is singularity free.
In the limit $L\rightarrow \infty$ we render the known results.
The new result in this use of piston is that the Casimir energy is
regularized explicitly and not just the force.

\section{Fermions-Bosons with Chiral Boundary Conditions}

Apart from the bag boundary conditions on the plates that can be used for
fermions, in the massless case we can use the so-called chiral boundary conditions. This stems from the
invariance of the massless Dirac equation
\begin{equation}\label{dir}
\mathcal{L}=\mathrm{i}\bar{\psi}\gamma^{\mu}\partial_{\mu}\psi
\end{equation}
under the transformation,
\begin{equation}\label{chir}
\psi'=\exp(\mathrm{i}\gamma_5\beta )\psi
\end{equation}
where $\beta$ is an arbitrary phase. The bag boundary conditions
\begin{equation}\label{mitbag1}
\mathrm{i}n^{\mu}\gamma_{\mu}\psi=\psi
\end{equation}
make the fermion current vanish outside the surface of the plates
\begin{equation}\label{mitbagcov1}
n^{\mu}\bar{\psi}\gamma_{\mu}\psi=0
\end{equation}
Although relation (\ref{mitbagcov1}) is chirally invariant, the
boundary condition (\ref{mitbag1}) is not \cite{ravdal1}.
The restoration of the chiral symmetry in the boundary conditions
can be achieved through the so called chiral boundary conditions,
\begin{equation}\label{chiralbc}
\mathrm{i}n^{\mu}\gamma_{\mu}\psi=\exp(\mathrm{i}\gamma_5\phi)\psi
\end{equation}
These conditions are invariant under the transformation (\ref{chir}) with $\phi$,
being promoted to a dynamical variable, transforming as $\phi\rightarrow \phi-2\beta$.

\begin{figure}[!t]
\centering
\includegraphics[scale=0.5]{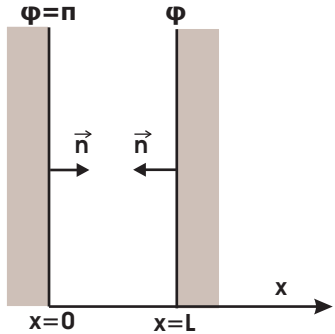}
\caption{The chiral boundary conditions}
\end{figure}

\noindent Consider a chamber made from two parallel plates
containing one vector boson gauge field satisfying perfect
conductor boundary conditions on the plate and one Majorana
fermion satisfying chiral boundary conditions on the plates.
Regarding the chiral boundary conditions the $\phi$ parameter is
considered to be $\phi=\pi$ on the plate located at $x=0$ and a
general value $\phi$ on the other plate on $x=\alpha$ (with $0\leq
\phi \leq 2\pi$). The reason to consider this setup is to examine
the behavior of the Casimir energy and of the Casimir force as a
function of the chiral parameter $\phi$. Moreover we want to
examine the minimum of the energy as a function of $\phi$. It is
known from reference \cite{ravdal1}, that the minimum
energy for a fermion with chiral boundaries is obtained for $\phi
=\pi$. We want to see how this result changes when we consider
bosons and fermions together, in a ``supersymmetric'' way.
(keeping in mind that chiral boundaries break supersymmetry.
The motivation using
bosons and fermions is as before, the cancellation of the
singularities in the Casimir energy and consequently in the
Casimir force, without the introduction of counter-terms to cancel
the singularities.

\noindent The vector boson contribution to the total Casimir
energy per unit area in the chamber is,
\begin{equation}\label{19}
\mathcal{E}_b(\alpha)=-\frac{\pi^2}{720\alpha^3}+\frac{3a}{\pi^2\tau^4}
\end{equation}
The fermionic Casimir energy per unit area with chiral boundaries
reads,
\begin{equation}\label{20}
\mathcal{E}_f(\alpha)=\frac{\pi^2}{720\alpha^3}-\frac{\phi^2}{384\pi^2\alpha^3}(\phi-2\pi)^2-\frac{3a}{\pi^2\tau^4}
\end{equation}
Adding the two bosons and one fermion contributions we obtain the
total Casimir energy in the chamber,
\begin{equation}\label{21}
\mathcal{E}_{\mathrm{total}}=-\frac{\phi^2}{384\pi^2\alpha^3}(\phi-2\pi)^2
\end{equation}
which is regular since the singularities of the two cancel each
other. From the above we easily obtain the Casimir force between
the plates,
\begin{equation}\label{22}
F_{\mathrm{total}}=-\frac{\phi^2}{64\pi^2\alpha^4}(\phi-2\pi)^2
\end{equation}
Thus we see that in the Casimir chamber with chiral boundaries the
total Casimir energy is regularized.

\subsection{Casimir Piston with Chiral Boundaries}

To complete the study of the chiral boundaries Majorana
fermion-vector boson chamber, we shall consider a piston using the
above combination of fields and boundary conditions. Particularly
suppose that we have a piston filled in the left chamber with a
Majorana fermion obeying chiral boundary conditions on $x=0$ and
$x=a$ (with $\phi =\pi$ on $x=0$ and general $\phi$ on $x=a$) and
with a vector boson in the other chamber with perfect conductor
boundary conditions at $x=a$ and $x=L$. Using the same procedure
as in the piston configuration of section 4, we obtain the total
Casimir energy, which is finite (free of singularities) and is
equal to,
\begin{equation}\label{21a}
\mathcal{E}_{\mathrm{total}}=-\frac{\pi^2}{720(L-\alpha)^3}+\frac{\pi^2}{720\alpha^3}-\frac{\phi^2}{384\pi^2\alpha^3}(\phi-2\pi)^2
\end{equation}
while the force on the plate at $x=a$ is given by
\begin{equation}\label{23}
F_{\mathrm{total}}=\frac{\pi^2}{480\alpha^4}+\frac{\pi^2}{120(L-\alpha)^4}-\frac{\phi^2}{128\pi^2\alpha^4}(\phi-2\pi)^2
\end{equation}
Notice that in the limit $L\rightarrow \infty$ the energy and the
force become equal to that of a single Majorana fermion between
plates with chiral boundary conditions.

\noindent A similar analysis holds for the case the boson and
fermion are put into opposite chambers. In that case the energy is
again finite of course and so is the force, which is equal to
\begin{equation}\label{24}
F_{\mathrm{total}}=-\frac{\pi^2}{120\alpha^4}-\frac{\pi^2}{480(L-\alpha)^4}+\frac{\phi^2}{128\pi^2(L-\alpha)^4}(\phi-2\pi)^2
\end{equation}
As before, the limit $L\rightarrow \infty$ renders the known
result of vector boson force between two plates. Let us note that,
as in the previous section, the regularization in the piston case
occurs at the energy density per unit area.

\section{Fermions with Bag Boundaries and Bosons with Robin Boundary Conditions}

We discussed in the previous sections the case having massless
Majorana fermions and vector bosons exist between parallel plates.
The number of the fields we used was determined by the way they
should appear in a vector supermultiplet of a $N=1$ supersymmetry.
Of course supersymmetry was broken by the boundary conditions
satisfied by fermions and bosons (bag or chiral for fermions and
perfect conductor for bosons). Although supersymmetry was broken,
the effect of using a vector boson and one Majorana fermion was
that the singularities in the total Casimir energy per unit area
(for the chamber) or in the Casimir energy density per unit area
(for the piston case) where cancelled. In this section we shall
consider a similar interplay between the number of fermions and
bosons using two equal mass scalar bosons  and a
massive Dirac fermion. The boundary conditions will be bag for
fermions and specific Robin boundary conditions for the scalar
bosons.

\noindent To start with, consider a fermion between two parallel
plates located at $x_d=0$ and $x_d=L$ embedded in a flat
$d$-dimensional spacetime. The Casimir energy for the fermion
equals to \cite{bag1,elizaldeslab}
\begin{equation}\label{25}
\mathcal{E}_f=-C(d)\int
\sum_{n}\frac{\mathrm{d}p^{d-2}}{(2\pi)^{d-2}}\sqrt{p^2+z_n^2+m_f^2}
\end{equation}
with $C(d)=2^{(d-1)/2}$ for $d$ odd and $C(d)=2^{(d-2)/2}$ for $d$
even. When the fermion obeys bag boundary conditions in the two
plates, the $z_n$ appearing above are solutions of the equation
\begin{equation}\label{26}
F(z)=0
\end{equation}
with
\begin{equation}\label{27}
F(z)=mL \sin(z)+z\cos(z)
\end{equation}
The roots of the above equation solve the eigenvalue problem for
the massive slab bag fermion \cite{bag1,elizaldeslab}.

\noindent Now consider a boson between two parallel plates located
at $x_d=0$ and $x_d=L$ obeying robin boundary conditions
\cite{ElizaldePistons,saharian1} of the form
\begin{equation}\label{28}
(1+\beta_1\partial_x)\phi (x,t)=0
\end{equation}
on the first plate (at $x=0$) and
\begin{equation}\label{29}
(1-\beta_2\partial_x)\phi (x,t)=0
\end{equation}
on the second plate ($\beta_{1,2}$ are arbitrary constants). Robin boundary conditions are known to
provide conformal invariance for field theories between parallel
plates \cite{saharian1} and are frequently used in the calculation
of the Casimir energy (see \cite{saharian1,ElizaldePistons} and
references therein). The interest in Robin boundary conditions comes from
the fact that there is a region in the parameter space for which
the Casimir forces are repulsive for small distances and
attractive for large distances. Thus stabilization of the distance
between the plates can be achieved.

\noindent We shall use Robin boundaries for the two massive scalar
bosons between two parallel plates in a $d$ dimensional spacetime
(the same setup as that of the fermion case). The Casimir energy for
the system with these boundary conditions is obtained by solving
the equation
\begin{equation}\label{30}
F(y)=(1-b_1b_2y^2)\sin y-(b_1+b_2)y\cos y=0
\end{equation}
with $b_i=\beta_i /L $. The above equation gives solution to the
spectral problem under Robin boundary conditions
\cite{ElizaldePistons,saharian1}. We shall use Dirichlet boundary
conditions in the $x=0$ boundary (this means that $b_1=0$) and
Robin boundary conditions on the $x=L$ boundary specified by,
\begin{equation}\label{31}
b_2=-\frac{1}{mL}
\end{equation}
Thus equation (\ref{30}) reads
\begin{equation}\label{32}
F(y)=mL \sin y+y\cos y=0
\end{equation}
The solutions of this equation, namely the roots $y_n$, solves the
spectral problem of the boson field between plates with boundary
conditions (\ref{28}) and (\ref{29}). The bosonic Casimir energy
of a massive boson reads
\begin{equation}\label{33}
\mathcal{E}_b=\int
\sum_{n}\frac{\mathrm{d}p^{d-2}}{(2\pi)^{d-2}}\sqrt{p^2+y_n^2+m_b^2}
\end{equation}
Notice that equations (\ref{32}) and (\ref{27}) are identical.
Thus the total Casimir energy for the two bosons and one fermion
field equals to
\begin{equation}\label{totcasen}
\mathcal{E}_{tot}=2\mathcal{E}_b+\mathcal{E}_f=2\int
\sum_{n}\frac{\mathrm{d}p^{d-2}}{(2\pi)^{d-2}}\sqrt{p^2+y_n^2+m_b^2}-C(d)\int
\sum_{n}\frac{\mathrm{d}p^{d-2}}{(2\pi)^{d-2}}\sqrt{p^2+z_n^2+m_f^2}
\end{equation}
The motivation to consider this fermion-boson
``supersymmetric" configuration ($N=1$ chiral supermultiplet) is
the observation that for $d=4$ and for $d=3$ the total Casimir
energy vanishes when the fermion and bosons have the same mass,
that is when $m_f=m_b$. This result holds only for these two $d$
values. Indeed, for $d=4$, the parameter $C(d)=2^{(d-2)/2}$ which
holds for $d=$ even becomes $C(4)=2$. The same occurs for $d=3$ on
the other relation. Thus although supersymmetry is broken on the
boundaries, the total Casimir energy vanishes. This result is
particularly interesting because it occurs only for $d=4$ when
$d=$ even and $d=3$ when $d=$ odd. This effect is known to occur
only in supersymmetric theories (see for example reference
\cite{malakoiapones}).

\section{Scalar Boson Piston with Neumann-Dirichlet-Dirichlet\\ Boundary Conditions}

In this section we shall study a chamber filled with two scalar
fields having the same mass. Although this differs only slightly from
the case studied in the previous section, the motivation is to show
another way to cancel the singularities from scalar bosons Casimir
energies between two parallel plates. In order to see this,
consider two scalar bosons in the space between two parallel
plates located at $x=0$ and $x=L$. One of them is assumed to
satisfy Dirichlet boundary conditions on both plates, while the
other one satisfies Neumann at $x=0$ and Dirichlet on the other
plate. Within the dimensional regularization method, the Casimir
energy between the two plates of the Neumann-Dirichlet boson is
equal to (in the end we shall put $d=4$ and $s=-1/2$)
\begin{align}\label{mixed}
&\mathcal{E}_{ND}=\frac{1}{(2\pi)^{d-2}}\pi^{\frac{d-2}{2}}
\frac{\Gamma(s-\frac{d-2}{2})}{\Gamma(s)}\sum_{n=0}^{\infty}\Big{[}\frac{\pi^2}{L^2}(n+\frac{1}{2})^2+m^2\Big{]}^{\frac{d-2}{2}-s}.
\end{align}
We shall use zeta-function regularization in order to compute the
above. Making use of the following relation \cite{elizalde}
\begin{align}\label{epsteinzhurwitzeta}
&\sum_{n=0}^{\infty}\Big{(}a(n+\frac{1}{2})^2+q\Big{)}^{-s}=
\\& \notag -\sqrt{\frac{\pi}{a}}{\,}\frac{\Gamma(s-\frac{1}{2})}{\Gamma
(s)}\frac{q^{-s+1/2}}{2}+\frac{2\pi^{s}q^{-s/2+1/4}a^{-s/2-1/4}}{\Gamma
(s)}\sum_{n=1}^{\infty}(-1)^nn^{s-1/2}K_{s-1/2}(2\pi
n\sqrt{\frac{q}{a}}).
\end{align}
the Casimir energy for the Neumann-Dirichlet scalar boson is
\begin{align}\label{twisted}
&\mathcal{E}_{ND}=-\frac{1}{2^{d-1}\pi^{(d-1)/2}}\frac{\Gamma(s-\frac{d-1}{2})}{\Gamma(s)}(m^2)^{-s+\frac{1}{2}+\frac{d-2}{2}}L
\\& \notag +\frac{2\pi^{s-(d-2)}}{\Gamma(s)}m^{\frac{1}{2}-(s-\frac{d-2}{2})}(\frac{\pi^2}{L^2})^{\frac{1}{4}-\frac{1}{2}(s-\frac{d-2}{2})}
\\& \notag \times
(2mL)^{-s+\frac{d-1}{2}}\sum_{n=1}^{\infty}(-1)^n(2mLn)^{s-\frac{d-1}{2}}K_{s-\frac{d-1}{2}}(2mnL)
\end{align}
with
\begin{align}\label{79}
&\sum_{n=1}^{\infty}(-1)^n(2mLn)^{s-\frac{d-1}{2}}K_{s-\frac{d-1}{2}}(2mnL)=
\\&
\notag\sum_{n=1}^{\infty}(4mLn)^{s-\frac{d-1}{2}}K_{s-\frac{d-1}{2}}(4mnL)-\sum_{n=1}^{\infty}(2mLn)^{s-\frac{d-1}{2}}K_{s-\frac{d-1}{2}}(2mnL)
\end{align}
In the above, the following relation was used.
\begin{equation}\label{sumtrick}
\sum_{q=1}^{\infty}(-1)^qf(r)=2\sum_{q=1}^{\infty}f(2r)-\sum_{q=1}^{\infty}f(r)
\end{equation}
In the same way we obtain the Casimir energy of the
Dirichlet-Dirichlet boson which is equal to
\begin{align}\label{untwisted}
&\mathcal{E}_{DD}=-\frac{1}{2^{d-1}\pi^{(d-2)/2}}\frac{\Gamma(s-\frac{d-2}{2})}{\Gamma(s)}m^2
\\& \notag +\frac{1}{2^{d-1}\pi^{(d-1)/2}}\frac{\Gamma(s-\frac{d-1}{2})}{\Gamma(s)}(m^2)^{-s+\frac{1}{2}+\frac{d-2}{2}}L
\\& \notag +\frac{1}{2^{d-3}}\frac{\pi^{s-(d-2)}}{\Gamma(s)}m^{\frac{1}{2}-(s-\frac{d-2}{2})}(\frac{L^2}{\pi^2})^{\frac{1}{4}+\frac{1}{2}(s-\frac{d-2}{2})}
\\& \notag \times \sum_{n=1}^{\infty}n^{s-\frac{d-1}{2}}K_{s-\frac{d-1}{2}}(2mnL)
\end{align}
where we have made use of the following analytic continuation
\cite{elizalde}
\begin{align}\label{epsteinzhurwitzeta1}
\zeta_{EH}(s;p)=-\frac{p^{-s}}{2}+\frac{\sqrt{\pi}{\,}{\,}\Gamma
(s-1/2)}{2\Gamma (s)}p^{-s+1/2}+\frac{2\pi^{s}p^{-s/2+1/4}}{\Gamma
(s)}\sum_{n=1}^{\infty}n^{s-1/2}K_{s-1/2}(2\pi n\sqrt{p}).
\end{align}
Thus the total Casimir energy between the plates is
$\mathcal{E}_{tot}=\mathcal{E_{DD}}+\mathcal{E_{ND}}$. As can be
easily seen, relations (\ref{untwisted}) and (\ref{twisted})
contain singularities due to the Gamma function
$\Gamma(s-\frac{d-1}{2})$ for $d=4$ and $s=-1/2$
(first line of (\ref{twisted}) and second line of (\ref{untwisted})).
It is obvious that the two
singular terms cancel each other, thus the final result of the
total Casimir energy is finite. The final
expression is equal to,
\begin{align}\label{totaltwisteduntwisted}
&\mathcal{E}_{tot}=\frac{2\pi^{s-(d-2)}}{\Gamma(s)}m^{\frac{1}{2}-(s-\frac{d-2}{2})}(\frac{\pi^2}{L^2})^{\frac{1}{4}-\frac{1}{2}(s-\frac{d-2}{2})}
\\& \notag \times
\Big{(}\sum_{n=1}^{\infty}(4mLn)^{s-\frac{d-1}{2}}K_{s-\frac{d-1}{2}}(4mnL)-\sum_{n=1}^{\infty}(2mLn)^{s-\frac{d-1}{2}}K_{s-\frac{d-1}{2}}(2mnL)\Big{)}
\\ & \notag
-\frac{1}{2^{d-1}\pi^{(d-2)/2}}\frac{\Gamma(s-\frac{d-2}{2})}{\Gamma(s)}m^2
\\& \notag +\frac{1}{2^{d-3}}\frac{\pi^{s-(d-2)}}{\Gamma(s)}m^{\frac{1}{2}-(s-\frac{d-2}{2})}(\frac{L^2}{\pi^2})^{\frac{1}{4}+\frac{1}{2}(s-\frac{d-2}{2})}
\\& \notag \times \sum_{n=1}^{\infty}n^{s-\frac{d-1}{2}}K_{s-\frac{d-1}{2}}(2mnL)
\end{align}
We shall not pursuit this further since we just wanted to show how
the cancellation of singularities within this setup occurs in four
dimensions.

\section{Conclusions}

We have studied the Casimir effect of a massless Majorana fermion
field and of a massless vector boson field confined between
parallel plates. Particularly we considered the vector boson and
the fermion field contained in a chamber and in a piston. We
used various configurations which we now briefly discuss. The way
the fields where chosen was according to a $N=1$ supersymmetric
Lagrangian that was actually broken due to the boundary conditions
obeyed by the fields on the parallel plates. The fermion field was
supposed to obey bag or chiral bag boundary conditions on the
plates while the vector boson field obeyed perfect conductor
boundary conditions.

\noindent We saw that when we use both massless Majorana fermions
and vector bosons in a chamber, both the total Casimir energy
density and the Casimir energy density per unit area is
singularity free. We used the cutoff method in order to explicitly
see this cancellation. Thus although supersymmetry was explicitly
broken on the boundary of the parallel plate chamber, a remnant of
supersymmetry was responsible for the cancellation of the
singularities.

\noindent Another interesting configuration we used in this
article was the piston, which consists of three parallel plates
placed at $x=0$, $x=a$ and $x=L$. We have started by putting in the $a$ chamber
the massless Majorana fermion and in the $L-a$ chamber  the
massless vector boson obeying bag and perfect conductor boundary
conditions respectively. We concluded that the energy density per unit area of
the system is regularized when we add the energy density
contributions from the two chambers of the piston. Then
integrating on the finite dimension we obtain the total energy  per unit area for
the piston. The limit $L\rightarrow \infty$ yields
the fermionic Casimir energy density for a Majorana fermion. The
Casimir force was calculated and in the limit $L\rightarrow
\infty$ turns to be identical with the fermion Casimir force between two
plates at a distance $a$.

\noindent Equally interesting is the case for which the vector
boson is put in the $a$ chamber and the Majorana fermion on the
$L-a$ chamber. We concluded that the energy density per unit area
is regularized. As before we calculated the total Casimir energy
and force which we found it to be finite. The limit $L\rightarrow
\infty$ yield the Casimir force and Casimir energy for a vector
boson between two parallel plates. We should mention that
this method yields a regularized result at the energy density level and
not just in the force level.

\noindent The interest in calculating massless vector boson
Casimir energies is obvious, since the photon belongs to that category.
Thus the confinement of the electromagnetic field in a slab leads
to a quantum effect, a vacuum energy density. We regularized the
vacuum energy density using a massless Majorana fermion in the
other chamber. That Majorana particle could be the photino.

\noindent The calculation of Majorana fermion Casimir energy
densities is well motivated. Majorana fermions occur quite
frequently in various research areas of physics. For example in
some models of universal extra dimensions, a WIMP candidate
(Weakly Interacting Massive Particle) is the Kaluza-Klein neutrino
(the universal extra dimensions Kaluza-Klein excitation) which is
a Majorana fermion \cite{vergadosme}. Actually the Dirac neutrino
is experimentally ruled out since the Z-induced neutron coherent
contribution would be too large, which does not happen in the
Majorana case. Another interesting appearance of Majorana fermion
states comes from the physics of superconductors and superfluids.
Particularly a Majorana bound state is theoretically predicted in
rotating superfluid $ ^3\mathrm{He-A}$ between parallel plates. In
the parallel plate geometry, the gap is about $10\mu$m and a
Magnetic field is applied along the plates. The Majorana vortex
bound state is associated with a singular vortex in chiral p-wave
superfluid \cite{superfluid}. Similar studies show that a Majorana
bound state exists in triplet superconductors (see \cite{triplet}).

\noindent Concluding on the Majorana-vector boson case, the
chamber and the piston setups where extended to the case of a
fermion field obeying chiral bag boundary conditions and we have found
that similar results hold.

\noindent Following the same supersymmetric recipe, regarding the
number of fields (of course supersymmetry is broken on the
boundaries), we studied a massive Dirac fermion and two massive
scalar bosons between two parallel plates. The spacetime dimensions is $d$ and the fermions obey bag
boundary conditions on the plates, while the scalar bosons obey
specific Robin boundary conditions. We found that for $d=4$ (for
$d=$ even) and for $d=3$ (for $d=$ odd) when the fermions and bosons
have the same mass, the total Casimir energy vanishes. It is
interesting to note that this occurs only for $d=3,4$.

\noindent Finally we presented another case in which the
singularities of the Casimir energy cancel. It is Casimir chamber
filled with two massive scalar bosons obeying Neumann and
Dirichlet boundary conditions at the plates. We found that under
these circumstances the Casimir energy of the system is
singularity free.

\noindent Regarding the Majorana-vector boson case, it would be
interesting to include finite temperature corrections or constant
magnetic field in the calculation of the Casimir energy. This is
of particular importance since these situations occur in
nano-devices and in other technological applications where
Majorana fermions appear \cite{superfluid}.

\end{document}